\documentclass[fleqn,twoside]{article}
\usepackage{amsmath,amssymb,bm}
\usepackage{espcrc2}

\usepackage{graphicx}
\usepackage[figuresright]{rotating}

\title{Analytic progress on exact lattice chiral symmetry}

\author{Y. Kikukawa\address[]{Department of Physics, Nagoya University,
        Nagoya 464-8602, Japan }
}       
\begin{document}

\begin{abstract}
Theoretical issues of exact chiral symmetry on the lattice are
discussed and related recent works are reviewed. 
For chiral theories, 
the construction with exact gauge invariance is reconsidered
from the point of view of domain wall fermion. The issue
in the construction of electroweak theory is also discussed.
For vector-like theories, 
we discuss 
unitarity (positivity), Hamiltonian approach, and
several generalizations 
of the Ginsparg-Wilson relation (algebraic and odd-dimensional).
%\vspace{1pc}
\end{abstract}

% typeset front matter (including abstract)
\maketitle

\section{Introduction 
%\\ 
%-- What are the remaining issues? --
}
\label{sec:introduction}

Three years has passed since the re-discovery of 
the Ginsparg-Wilson relation\cite{ginsparg-wilson-rel}. 
Now we know how to construct a 
{\it gauge-covariant} and {\it local} lattice Dirac operator which satisfies
the Ginsparg-Wilson relation\cite{fp-action-D,overlap-D,locality-of-overlap-D}.
Lattice fermion theories with such a Dirac operator
nicely reproduce 
the properties of massless Dirac fermion:
the exact symmetry of fermion action\cite{exact-chiral-symmetry}, 
the chiral anomaly from functional 
measure\cite{exact-chiral-symmetry,kikukawa-yamada,adams,fujikawa,suzuki,chiu},  
and the novel index theorem on the 
lattice\cite{index-theorem-on-lattice,overlap}.
Moreover, it has opened the possibility to construct chiral 
gauge theories on the lattice with exact gauge
invariance\cite{abelian-chiral-gauge-theory,nonabelian-chiral-gauge-theory,effective-action-abelian-chiral}.

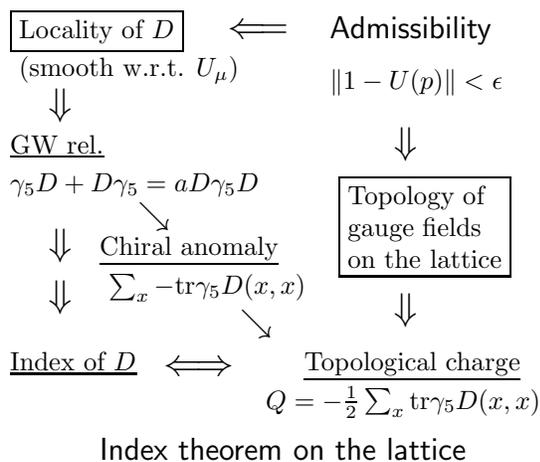
\begin{figure}[h]
  \begin{center}
{\unitlength 0.34mm
\begin{picture}(80,160)

\put (-60,145){{\fbox{Locality of $D$}} }
\put (-60,130){ (smooth w.r.t. $U_\mu$) }

\put(-45,115){\Large $\Downarrow$}

\put (-60,100){{\underline{GW rel.}} }
\put (-60,85){$\gamma_5 D + D\gamma_5 =a D \gamma_5 D$ }

\put(-10,72){\large $\searrow$}
\put(30,30){\large $\searrow$}

\put (-25,60){{\underline{Chiral anomaly}} }
\put (-25,45){ $\sum_x -{\rm tr}\gamma_5 D(x,x)$}

%\put(-55,40){\Large $\Downarrow$}
\put(-45,60){\Large $\Downarrow$}
\put(-45,40){\Large $\Downarrow$}

\put (-60,15){{\underline{Index of $D$}} }

\put (0,+13){{\Large $\Longleftrightarrow$}}

\put(25,145){\Large $\Longleftarrow$}

\put (+65,145){\large\textsf{Admissibility}}
\put(+65,125){$ \left\Arrowvert  1- U(p) \right\Arrowvert < \epsilon$}

\put(90,100){\Large $\Downarrow$ }

%\put(+50,55){{ \fbox{\vbox{\hbox{Topology of}\hbox{lattice gauge fields}}}}}
\put(+65,55){{ \fbox{\vbox{\hbox{Topology of}\hbox{gauge fields}
                           \hbox{on the lattice}}}}}

\put(90,35){\Large $\Downarrow$ }

\put(+55,15){{\underline{Topological charge}}}
\put(+40,0){$Q=-\frac{1}{2}\sum_x {\rm tr}\gamma_5 D(x,x)$}
%\put (+40,-30){ (smooth w.r.t. $U_\mu$) }

\put(-25,-20){\large
\underline{\textsf{Index theorem on the lattice}}}

\end{picture}
}
\caption{Structure of Ginsparg-Wilson fermions}
\label{fig:GW-fermion}
  \end{center}
\end{figure}

Let me first recall the basic structure of the Ginsparg-Wilson fermions, 
which may be summarized as in the figure~\ref{fig:GW-fermion}.
For the exact chiral symmetry based on the Ginsparg-Wilson relation
%\begin{equation}
%  D \gamma_5 + \gamma_5 D = a D \gamma_5 D
%\end{equation}
to make sense, the locality of the lattice Dirac operator is crucial.
Otherwise the lattice chiral transformation, defined as follows
\begin{equation}
  \delta \psi(x) = \gamma_5(1- a D) \psi(x), \quad
  \delta \bar \psi(x) = \bar \psi(x) \gamma_5,
\end{equation}
cannot be regarded to be a local transformation.
The locality has been proved rigorously 
for Neuberger's (overlap) Dirac operator under the so-called 
admissibility condition\cite{locality-of-overlap-D,neuberger-bound}
on the the plaquette variable, $U(p)$:
\begin{equation}
\parallel 1- U(p) \parallel \le \epsilon, \quad
\epsilon < \frac{1}{30} \ \ (m_0=1).
\end{equation}
The admissibility condition has important dual roles.
On one side, it ensures the locality and the smooth dependence
of lattice Dirac operators with respect to gauge fields.
On the other hand, it gives rise to 
the topological structure in the space of lattice gauge fields.
Both are crucial for the index theorem to hold on the lattice
at a finite lattice spacing.

\vspace{-1.5em}

\vspace{-3em}
This structure of the Ginsparg-Wilson fermion tells us that
the question of the exact chiral symmetry on the lattice 
is intimately related to the
question about the behavior of lattice gauge fields.
In fact, in the construction of chiral gauge theories,
one needs to explore the topological structure of the
space of admissible gauge fields. 
It is needed 
for both the cohomological classification of 
the topological field in 4+2 dimensions which is 
related to the gauge anomaly cancellation, 
and the proof of the global integrability 
of the functional measure of Weyl fermions.
These issues, in particular for non-abelian cases,
turn out to be rather tough to address and 
one need to struggle with the admissibility condition.
In lattice QCD, on the other hand,
the gauge field action is the type of the Wilson action
\begin{equation}
S_G = \beta \sum_p \frac{1}{3} \, {\rm Re \, Tr}\left( 1- U(p) \right)
\end{equation}
and the admissibility condition is {\it not} imposed. 
Then it is necessary to understand how far one can maintain
chiral property, locality and also universality.
This is an urgent issue for numerical applications. It
will be covered 
in the talk (article) by Hern\'andez\cite{Pilar-review}.
%It is also an important and challenging issue to 
%seek lattice Dirac operators satisfying the Ginsparg-Wilson 
%relation with better locality properties. 
%It is desirable to have the systematic understanding about
%the conditions which 
%determines the locality properties of a lattice Dirac operator.
%I would like to to summarise the properties of these lattice Dirac operators.

In this talk (article), I would rather discuss
the issues in theoretical interests (so far).
First of all, I would like to discuss more on 
lattice chiral gauge theories. 
It is highly desirable
to extend L\"uscher's construction of abelian chiral gauge
theories with exact gauge invariance 
to non-abelian cases. Because of the reason I mentioned
above, however, little progress 
is obtained so far. 
Another interesting question 
is to seek a practical formulation of chiral gauge theories
which can be usable for non-perturbative studies through numerical methods.
Although chiral gauge theories are a difficult case for
numerical simulations, such a formulation can provide 
tools to get insights into the dynamics of chiral gauge theories.
I will discuss the use of domain wall fermion in this respect and
will argue that it provides a hint for a practical implementation.
Also I will discuss the issue in 
the construction of electroweak SU(2)$\times$U(1) gauge theory 
on the lattice. 

Secondly, I would like to discuss more on the basic properties of 
the Ginsparg-Wilson Dirac fermions. Unitarity is one of the 
basic properties which is not fully examined yet
and I will review recent results on this question.
There are several proposals for new types of lattice Dirac operators.
I will discuss an algebraic generalization of the Ginsparg-Wilson 
relation and its consequences.
Finally, I will review recent works on the Ginsparg-Wilson relation in
odd dimensions.

For notational simplicity the lattice spacing $a$ is set to unity
in the following.

\newlength{\minitwocolumn}
\setlength{\minitwocolumn}{0.5\textwidth}

\section{More on chiral gauge theories}
\label{sec:chiral-gauge-theory}

Once the lattice Dirac operator satisfying the Ginsparg-Wilson
relation is obtained, Weyl fermion and its functional measure
can be defined through the chiral projectors 
$\hat \gamma_5 =\gamma_5(1- D)$, $\gamma_5$ 
by introducing 
orthogonal chiral bases
$\left\{v_i\right\}$ and $\left\{\bar v_k\right\}$ as 
$\hat \gamma_5 v_i(x) = + v_i(x)$, 
$\bar v_k(x) \gamma_5 = - \bar v_k(x)$ .
Then the partition function(chiral determinant) of the Weyl fermion
can be evaluated as 
\begin{equation}
Z_{\rm W} = \det \left( \bar v_k, D v_j \right)   
          = \det \left( \bar v_k, v_j \right)   .
\end{equation}
When we adopt Neuberger's (overlap) Dirac operator,
\begin{equation}
D=\frac{1}{2}\left( 1+ \gamma_5 \frac{H}{\sqrt{H^2}} \right)
\end{equation}
where $H=\gamma_5(D_{\rm w}-m_0) (0 < m_0 < 2)$,
the above partition function 
reproduces the vacuum overlap formula
of chiral determinant\cite{overlap}.

However, there remains a certain ambiguity in the 
definition of the functional measure and
the partition function. In fact, when we choose a different
chiral basis, which should be related to the original one by
a unitary transformation,
$v_i(x) \rightarrow \tilde v_i(x) = \sum_j v_j(x) Q_{ji}$,
the functional measure is changed by the phase factor
$\det Q$.

L\"uscher proposed a method to fix the phase of the 
functional measure by imposing
the locality of the field equation,
the gauge-invariance, and the smooth dependence with
respect to gauge 
field\cite{abelian-chiral-gauge-theory,nonabelian-chiral-gauge-theory}. 
He gave a constructive proof that 
such a measure indeed exists in anomaly-free abelian chiral gauge theories.

\subsection{Use of domain wall fermion}

Since the vacuum overlap formula 
was originally derived from domain wall 
fermion\cite{domain-wall-fermion},
we may reconsider the above problem 
from the point of view of domain wall fermion.
For simplicity we adopt
the vector-like domain wall 
fermion\cite{boundary-fermion,boundary-fermion-QCD,blum-soni},
which is 
defined by the five-dimensional Wilson-Dirac fermion 
with a negative mass ($0< m_0 < 2$)
in a finite extent of the fifth-dimension,
$x_5=a_5 t$, $t \in [-N+1,N]$:
\begin{equation}
%S_{\rm DW}=
a_5 \sum_{t=-N+1}^N \sum_x \bar \psi(x,t) 
\left( D_{\rm 5w}
%-P_L \partial_t - P_R \partial_t^\dagger 
-m_0\right) \psi(x,t) .
\end{equation}
($a_5$ is the lattice spacing of
the fifth dimension.)
We first note that 
there is another possibility to couple gauge field
to the chiral zero modes of 
the vector-like domain wall fermion:
one may introduce a five-dimensional gauge field
which is varying in the fifth direction so that
the gauge field on the right wall, $U_\mu(x)$, 
is interpolated smoothly back to a trivial field, $U_\mu^0=1$, 
on the left wall,
$U_\mu(x,t)=\left(U_k(x,t),1\right) \ : \ U_k^0=1 \rightarrow U_k(x)$.
\hspace{1em}
\begin{minipage}[t]{\minitwocolumn}
{\unitlength 0.5mm
\begin{picture}(160,75)

\put(15,5){\line(1,0){80}}

% interpolation
{\linethickness{0.5mm}

\put(10,10){\line(1,0){30}}

\put(60,30){\line(1,0){35}}
\qbezier(40,10)(45,10)(50,20)
\qbezier(50,20)(55,30)(60,30)

\multiput(0,10)(5,0){4}{\line(1,0){3}}
\multiput(95,30)(5,0){4}{\line(1,0){3}}
}

%Walls
{\linethickness{0.3mm}
\multiput(-8,0)(80,0){2}{
  \begin{picture}(50,60)
    \put(20,5){\line(0,1){45}}
    \put(50,25){\line(0,1){45}}
    \put(20,5){\line(3,2){30}}
    \put(20,50){\line(3,2){30}}
  \end{picture}
}
}

\put(12,-2){$t=-N+1$}
\put(92,-2){$t=N$}

\put(20,40){$\psi_L(x)$}
\put(100,40){$\psi_R(x)$}

\put(17,25){$U^0_k =1$}
\put(97,20){$U_k(x)$}

\put(60,40){$ U_\mu(x,t)$}
%\put(65,35){$(- m_0 )$}
\put(0,-15){Figure 2. DWF for chial gauge theory}
\end{picture}
}
\end{minipage}
\vspace{2em}

\noindent Then, instead of treating 
the five-dimensional gauge field dynamically
(like the original proposal by Kaplan\cite{domain-wall-fermion}), 
one may consider to construct {\it a gauge-invariant and
local counter term, $C\left[U_\mu(x,t)\right]$, 
which can compensate the five-dimensional dependence
of the partition function of the domain wall fermion}:
\[
\det \left( D_{\rm 5w}-m_0 \right) \times {\rm e}^{C} :
\ {\rm functional \ of \ }U_\mu(x,N) .
\]
If such a counter term would exist, the domain wall fermion
could be reduced to a four-dimensional system which depends only on the
gauge field on the right boundary wall. 
This reduction could be achieved in a gauge-invariant and local manner.
This problem, as we will see below,
turns out to be equivalent to the problem addressed by L\"uscher, i.e. 
the question of the gauge invariant and local construction of the 
functional measure of the Weyl 
fermions\cite{nonabelian-chiral-gauge-theory,aoyama-kikukawa,chiral-domain-wall-fermion}. 

\vspace{.6em}
\noindent\textsf{\underline{Integrability condition for domain wall fermion}}
\vspace{.2em}

The dependence of the partition function 
of domain wall fermion on the five-dimensional gauge field
can be examined by considering a smooth variation of the 
five-dimensional gauge field.
We assume that the variation is supported
in the finite interval which is fixed in size w.r.t. $N$:
$\delta U_\mu(x,t) = \eta_\mu(x,t) U_\mu(x,t); 
\ t \in [-\Delta+1,\Delta]$.

\begin{minipage}[t]{\minitwocolumn}
{\unitlength 0.6mm
\begin{picture}(160,40)
\put(20,0){\line(1,0){85}}
\put(62.5,-1.5){\line(0,1){3}}
%\put(59.5,-6){0}

%% lattice 
\multiput(20,0)(5,0){18}{\circle*{1}}

\put(20,0){\line(0,1){30}}
\put(10,-6){\small $-N+1$} 
\put(105,0){\line(0,1){30}}
\put(105,-6){\small $N$}

% interpolation
{\linethickness{0.4mm}

\put(20,5){\line(1,0){30}}

\qbezier(50,5)(55,5)(60,15)
\qbezier(60,15)(65,25)(80,25)

\qbezier(49,5)(52,5)(55,15)
\qbezier(55,15)(60,25)(70,25)

\put(70,25){\line(1,0){35}}

}

\multiput(50,0)(0,5){6}{\line(0,1){3}}
\put(42,-6){$-\Delta+1$}
\multiput(75,0)(0,5){6}{\line(0,1){3}}
\put(73,-6){$\Delta$}

\put(22,8){$U^0_k(x)$}
\put(89,28){$U^1_k(x)$}
%\put(50,35){\underline{$U_\mu(x,t)$}}

\put(52,22){$c_2$}
\put(57,13){$\uparrow$}
\put(62,10){$c_1$}

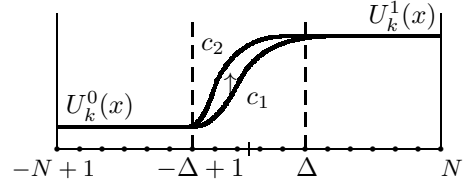
\put(10,-15){Figure 3. Interpolating 5D gauge field}
\end{picture}
}
    \label{fig:interpolation-in-fifth-dimension}
\end{minipage}
\vspace{2.5em}

\noindent Then we have
\begin{equation}
  \delta \det \left(D_{\rm 5w}-m_0 \right) 
 = \sum_t %{t=-\Delta+1}^\Delta 
\sum_x \eta_\mu^a(x,t) J_\mu^a(x,t),
\end{equation}
\begin{equation}
  J_\mu^a(x,t)=
{\rm Tr}\left( V_\mu^a(x,t) \frac{1}{D_{\rm 5w}-m_0}\right).
\end{equation}
A crucial point here is 
the locality property of the current $J_\mu^a(x,t)$ 
in the interpolation region $t\in[-\Delta+1,\Delta]$. 
It is possible to show rigorously that 
the current %$J_\mu^a(x,t)$ 
is a local functional of the five-dimensional
gauge field, provided that the gauge field 
satisfies the five-dimensional analog
of the admissibility condition:
\begin{equation}
\parallel 1- U_{\rm 5d}(p) \parallel \le \epsilon, \quad
\epsilon < \frac{1}{50} \ \ (m_0=1),
\end{equation}
where $U_{\rm 5d}(p)$ is the five-dimensional plaquette variable. 
The immediate consequence of this fact is that in the limit 
$N\rightarrow \infty$,  
the current $J_\mu^a(x,t)$ in the interpolation region 
does not actually depend on the specific choice of the boundary condition
for the domain wall fermion (Dirichlet boundary condition).
We may replace the current  by that of the
five-dimensional Wilson fermion 
subject to the anti-periodic boundary condition(AP)
in the fifth dimension which is enlarged 
twice as large as the original size (see the figure 4).
Because of the periodicity, 
the five-dimensional gauge field of the above configuration
can be regarded to define
a closed path, $c_1+(-c_0) : U_k^0 \rightarrow U_k^1 \rightarrow
U_k^0$ (or $c_2+(-c_0)$) in the space of four-dimensional gauge fields.
By noting 
the fact that the interpolation path $c_0$ can be chosen arbitrarily, 
and the reflection property of the five-dimensional Wilson
fermion under $  P: t \rightarrow -t +1 $;
$  U_k(x,t) \rightarrow U_k(x,-t+1)^{-1}$; 
$D_{\rm 5w} \rightarrow P \gamma_5 D_{\rm 5w}^\dagger \gamma_5 P$,

\begin{minipage}[t]{\minitwocolumn}
{\unitlength 0.45mm
\begin{picture}(160,60)

% base line
\put(10,0){\line(1,0){140}}
% lattice 
\multiput(10,0)(5,0){29}{\circle*{1}}

\put(0,-8){\small $-N+1$} 
\put(44.0,-8){\small 0}
\put(78.0,-8){\small $N$}
\put(112.5,-8){\small $2N$}
\put(148.0,-8){\small $3N$}

\multiput(35,0)(0,5){6}{\line(0,1){3}}
\multiput(55,0)(0,5){6}{\line(0,1){3}}

\multiput(105,0)(0,5){6}{\line(0,1){3}}
\multiput(125,0)(0,5){6}{\line(0,1){3}}

% interpolation
{\linethickness{0.5mm}

\put(10,5){\line(1,0){25}}

\qbezier(35,5)(40,5)(45,15)
\qbezier(45,15)(50,25)(55,25)

\qbezier(35,5)(37,5)(40,15)
\qbezier(40,15)(45,25)(55,25)

\put(55,25){\line(1,0){50}}

\qbezier(105,25)(110,25)(113,20)
\qbezier(113,20)(120,5)(125,5)

\put(125,5){\line(1,0){25}}

}

\put(10,8){$U^0_k(x)$}
\put(140,8){$U^0_k(x)$}
\put(75,28){$U^1_k(x)$}
%\put(75,10){$U_\mu(x,t)$}

\put(45,5){$c_1$}
\put(40,25){$c_2$}
\put(105,8){$-c_0$}

\put(40,45){$c_1,c_2$}
\put(58,45){\vector(1,0){22}}
\put(38,45){\line(-1,0){30}}

\put(8,43.5){\line(0,1){3}}
\put(79,43.5){\line(0,1){3}}
\put(150,43.5){\line(0,1){3}}

\put(107,45){$-c_0$}
\put(121,45){\vector(1,0){29}}
\put(105,45){\line(-1,0){30}}

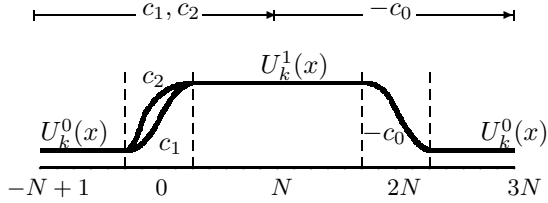
\put(10,-20){Figure 4. 5D gauge field for anti-periodic b.c.}
\end{picture}
}
\vspace{1em}
%    \label{fig:loop-in-fifth-dimension-II}
\end{minipage}
\vspace{2.5em}

\noindent
we can derive %another identity 
an identity for 
a subtracted partition function of
domain wall fermion: %defined by%, $\bar Z_{\rm DW}^{c}$, defined by
\begin{equation}
\label{eq:dwf-integrability}
  \ln \bar Z_{\rm DW}^{c_2}-  \ln \bar Z_{\rm DW}^{c_1}
= i   Q_{\rm 5w}^{c_2-c_1},
\end{equation}
\begin{equation}
\bar Z_{\rm DW}^{c} 
\equiv  
\lim_{N \rightarrow \infty}
\frac{\det \left(D_{\rm 5w}-m_0 \right)_c
}
{\left\vert 
\det \left(D_{\rm 5w}-m_0 \right)^{c-c}_{[AP]}
\right\vert^{\frac{1}{2}}}.
\end{equation}
$Q_{\rm 5w}^{c_2-c_1}$ is the complex phase of the 
determinant of five-dimensional Wilson fermion subject 
to anti-periodic b.c. and coupled to the five-dimensional
gauge field representing the loop $c_2-c_1$,
\begin{equation}
  Q_{\rm 5w}^{c_2-c_1}
\equiv \lim_{N \rightarrow \infty}
{\rm Im} \ln \det 
\left(D_{\rm 5w}-m_0 \right)^{c_2-c_1}_{[AP]}.
\end{equation}
This term is known to reproduce the Chern-Simons term
in the continuum limit\cite{so,coste-luscher}. Then we can conclude that 
the dependence on the five-dimensional interpolation
is governed by the lattice counter part of the Chern-Simons term.

\vspace{.6em}
\noindent\textsf{
\underline{Local cohomology problem in 5+1 dimensions}}
\vspace{.2em}

$Q_{\rm 5w}^{c_2-c_1}$ can always be expressed by 
a local field, if the closed path $c_2-c_1$ is contractible.
In fact, by introducing a continuous parameter $s \in [0,1]$
which parameterizes the deformation of the five-dimensional gauge field
$(U_k^0(x),1) \rightarrow U_\mu(x,t)$,
one obtains
\[
Q_{\rm 5w}^{c_2-c_1}
=   
\sum_{x,t}
\int_0^1 ds \, 
\left.
\left\{ 
\eta_\mu^a(x,t) \, {\rm Im} J_\mu^a(x,t)  \right\}\right\vert_{U_\mu(s)}.
\]
The local field in the r.h.s. is not gauge-invariant in general, but
there is a possibility to satisfy the gauge-invariance
by adding a certain total divergence term, which
does not affect $Q_{\rm 5w}^{c_2-c_1}$ itself:
\[
\delta_{\rm G}\left\{
\eta_\mu^a(x,t) \, {\rm Im} J_\mu^a(x,t) - \partial_\mu^\ast 
K_\mu(x,t) \vert_{U_\mu,\eta_\mu} 
\right\} = 0 .
\]

The above equation defines a local cohomology problem.
It can be re-formulated 
with the topological field in 
5+1 dimensions and 
can be solved non-perturbatively in abelian 
chiral gauge 
theories\cite{topology-on-the-lattice,noncomutative-brs-cohomology} and
in all orders of lattice perturbation theory in
non-abelian chiral gauge 
theories\cite{perturbation-theory,cohomology-non-abelian}. 
In these cases, one can show that 
there is the gauge-invariant and local density of 
$Q_{\rm 5w}^{c_2-c_1}$, 
{\it if the anomaly-free 
condition is satisfied} \cite{chiral-domain-wall-fermion}.

\vspace{.6em}
\noindent\textsf{
\underline{Reduction to four-dimensions and factorization}}
\vspace{.2em}

If $Q_{\rm 5w}^{c_2-c_1}$ can be written in terms of 
a gauge-invariant local density 
{\it for all possible closed path $c_2-c_1$ in the space of gauge fields},
then it is possible to write $Q_{\rm 5w}^{c_2-c_1}$ as 
\begin{equation}
 Q_{\rm 5w}^{c_2+(-c_1)}
=C_{\rm 5w}^{c_2}  -C_{\rm 5w}^{c_1},
\end{equation}
where 
$C_{\rm 5w}^{c_1}$ and $C_{\rm 5w}^{c_2}$ are gauge-invariant and
local terms 
associated with the interpolation paths $c_1$ and $c_2$, respectively.
Then from the identity Eq.~(\ref{eq:dwf-integrability}) it follows that 
\begin{equation}
  \bar Z_{\rm DW}^{c_1} \cdot e^{i C_{\rm 5w}^{c_1}}
=  \bar Z_{\rm DW}^{c_2} \cdot e^{i C_{\rm 5w}^{c_2}}.
\end{equation}
This result implies that
the partition function of the domain wall fermion
can be made independent of the interpolation path by
adding the local counter term.

We can work out explicitly the subtracted partition function with
the local counter term, 
$\bar Z_{\rm DW}^{c_1} \cdot e^{i C_{\rm 5w}^{c_1}}$,  
using the master formula 
in which 
the partition functions of the five-dimensional Wilson-Dirac 
fermions are expressed in terms of the transfer 
matrices\cite{truncated-overlap,luscher-dwf}.
It turns out that the partition function factorizes into 
the product of two chiral determinants in the overlap formula:
\begin{equation}
\bar Z_{\rm DW}^{c_1} \cdot e^{i C_{\rm 5w}^{c_1}}
=\det\left(\bar v_k,D v_j \right)
\cdot  \det\left(\bar v_k,D v_j^0 \right)^\ast ,
\end{equation}
where $D$ is Neuberger's (overlap) Dirac operator with the hermitian operator
$H$ defined through the transfer matrix $T= \exp(-a_5 H)$
of the five-dimensional Wilson-Dirac fermion, 
or more simply 
$H=\gamma_5(D_{\rm w}-m_0)/(1+a_5 (D_{\rm w}-m_0))
$\cite{borici}.
$\left\{v_j^0 \right\}$ and $\left\{v_j \right\}$ are
the chiral bases with respect to 
$U_k^0$ on the left wall and $U_k$ on the right wall, 
respectively.
Note, however, that in this case
the phase of the basis $\left\{v_j\right\}$ is fixed as follows:
\[
  v_i(x) = \left\{
       \begin{array}{ll}
          v_i^1(x) \, \,
e^{i \phi(c_1)} 
\, e^{-iC_{\rm 5w}(c_1)} %W(c_1)^{-1} 
& (i=1) \\
          v_i^1(x)            & (i \not = 1) 
       \end{array}
           \right. ,
\]
\[
e^{i \phi(c_1) }= 
\frac{ \scriptstyle
 \det \left( v_i^1, \biggl\{ 
%U_{5,\Delta}^{-1} 
\prod_{t=-\Delta+1}^{\Delta} T_t 
%U_{5,t-1}^{-1} 
    \biggr\} \,  v_j ^0\right)}
{\scriptstyle
\left\vert  \det \left( v_i^1, \biggl\{ 
%U_{5,\Delta}^{-1} 
\prod_{t=-\Delta+1}^{\Delta} T_t 
%U_{5,t-1}^{-1} 
    \biggr\} \,  v_j^0 \right)\right\vert} .
%\,
%\times
%{\scriptstyle
%\det\left(P_R + P_L \prod_{t=-\Delta+1}^{\Delta}U_{5,t} \right)}
\]
Then the chiral determinant 
$\det\left(\bar v_k,D v_j \right)$ is gauge-invariant
by construction !
We can check that the above choice of basis would define
the functional measure of Weyl fermions with 
the desired properties
in anomaly-free chiral gauge theories.
(See also \cite{adiabatic-phase-chioce} for 
a recent proposal how to fix the 
complex phase in the vacuum overlap formula.) 

In this manner, domain wall fermion can provide 
a gauge-invariant partition function of the Weyl fermions 
in four dimensions,
through the integrability condition for
domain wall fermion 
Eq.~(\ref{eq:dwf-integrability}) and the cohomological problem
in 5+1 dimensions.
The sufficient condition is that 
{\it there exists a gauge-invariant and local density 
of $Q_{\rm 5w}^{c_2-c_1}$(lattice Chern-Simons term)
for all possible closed path $c_2-c_1$ in the space of gauge fields}.

\subsection{A hint for practical implementation}

The fact that domain wall fermion (as a five-dimensional
Wilson-Dirac fermion) can provide
a concrete example of L\"uscher's construction
suggests that the continuous interpolation in the space of 
(admissible) lattice gauge fields can be discretized, 
{\it without loosing the topological properties of the
gauge anomaly of the Ginsparg-Wilson Weyl 
fermions}\cite{geometrical-aspect,nonabelian-chiral-gauge-theory}.
In fact, it is possible to define
the topological field {\it on six-dimensional lattice}
which can capture the global aspects
of the gauge anomaly. 

To see this, let us consider a two-dimensional
surface in the space of lattice gauge fields and
introduce lattice coordinates $t$ and $s$ on the surface. 
For each lattice site $(s,t)$, we associate 
a four-dimensional lattice gauge field and 
a chiral basis with this gauge field:
$\{v_j(x)^{(s,t)} \} ; U_k(x,s,t)$. 
Then we can define a two-dimensional U(1) gauge field
($G_s,G_t$) on the lattice as follows:
\begin{equation}
G_s(s,t) \equiv 
\frac{
\det\left(v_i^{(s+1,t)},v_j^{(s,t)}\right) 
}
{\vert 
\det\left(v_i^{(s+1,t)},v_j^{(s,t)}\right) 
\vert } 
\end{equation}
and a similar definition for $G_t(s,t)$. The change of the basis
by a unitary transformation at each site induces U(1) lattice gauge
transformation. 

Then we may consider the plaquette variable 
of the U(1) gauge field given by
\[
U(s,t)\equiv
G_s(s,t)G_t(s+1,t)G_s(s,t+1)^{-1}G_t(s,t)^{-1}
\]
and may define a topological term 
\begin{equation}
  \sum_{s,t} \frac{1}{2 \pi i} \ln U(s,t).
\end{equation}
One can show that this topological term reduces 
in the continuum limit to
L\"uscher's topological field in 4+2 
dimensions\cite{nonabelian-chiral-gauge-theory}. 
Therefore, as long as the two-dimensional lattice is fine enough, 
the topological term can capture the global aspect of the
gauge anomaly of the Weyl fermion on the lattice.
We may also consider a Wilson loop
of the U(1) gauge field given by $\prod_{loop} G_i(s,t)$.
It turns out to be identical to 
the lattice Chern-Simon term $\exp(i Q_{\rm 5w}^{\rm loop})$
in a certain limit.
Thus we observe the structure over four-, five-, and six-dimensional
lattices similar to the descent relation. 
%\cite{descent-relation-zumino}.

The topological field on the six-dimensional lattice
may be used to solve the cohomology problem and
to construct the gauge-invariant and local density 
of the Chern-Simon term. % explicitly.
Since the interpolation is discrete, we may
perform explicitly the check of the global integrability condition.

\subsection{Electroweak theory}

It is highly desirable
to extend L\"uscher's construction of abelian chiral gauge
theories with exact gauge invariance 
to non-abelian cases. As a first step towards this direction,
one may consider to extend the gauge-invariant construction to
electroweak SU(2)$\times$U(1) gauge 
theory\cite{kikukawa-nakayama-suzuki}.

Electroweak theory is the chiral gauge theory 
of left-handed leptons and quarks in SU(2) doublet
and right-handed quarks in SU(2) singlet. Taking into account
of the color degrees of freedom, there are four doublets 
in each generation.
The space of the admissible SU(2)$\times$U(1) gauge
fields is divided into the topological 
sectors\cite{abelian-chiral-gauge-theory,ML-topology}, 
each one is the product of a U(1) topological sector 
and a SU(2) topological sector.

In the vacuum sector of the U(1) gauge fields,
where 
any configuration can be deformed smoothly to the trivial one 
$U^{(1)}(x,\mu)=1$,
we can turn off the hypercharge gauge coupling. Then
the theory can be regarded as 
vector-like due to the pseudo reality of SU(2).
It is indeed possible
to make the fermion measure 
defined globally
for all topological sectors of 
SU(2) by the following choice of the basis for a pair of 
doublets (a,b)\cite{abelian-chiral-gauge-theory}:
\begin{eqnarray}
\label{eq:su2-basis-global-a}
w_j^{(a)}(x) &=& u_j(x),   \\
\label{eq:su2-basis-global-b}
w_j^{(b)}(x) &=& \left( \gamma_5 C^{-1} \otimes 
i \sigma_2 \right) \, \left[ u_j(x) \right]^\ast ,
\end{eqnarray}
where $\hat \gamma_5 u_j(x) = - u_j(x)$. 
It is nothing but the %One may adopt the so-called 
symplectic basis for real representations 
considered by Suzuki\cite{real-rep-suzuki}.
This fact 
implies the cancellation of Witten's SU(2) anomaly 
(cf. \cite{neuberger-su2,OB-IC}).
Given the basis for the SU(2) doublets defined globally, 
one may try to extend the fermion measure 
to incorporate the U(1) gauge field
following the reconstruction 
theorem\cite{nonabelian-chiral-gauge-theory}.

The issues in this approach are 
the local cohomology problem
and the proof of the global integrability condition.
Fortunately, the cohomology problem can be solved 
for the U(1) part by the trick to treat the SU(2) gauge field as a 
background\cite{YK-YN-anomaly,abelian-chiral-gauge-theory}.
As to the global integrability condition, 
it is proved
for ``gauge loops'' in the space of the %admissible 
U(1) gauge fields
with arbitrary SU(2) gauge field in the background.
For ``non-gauge loops'', however, 
the proof is given so far 
only for the classical SU(2) instanton 
backgrounds\cite{kikukawa-nakayama-suzuki}.
To proceed, it seems that the information 
on the topological structure of the admissible SU(2)
gauge field is required.

\section{More on Ginsparg-Wilson fermions}
\label{sec:GW-Dirac-fermion}

\subsection{Unitarity and Hamiltonian approach}

Unitarity is one of the remaining issues 
about the Ginsparg-Wilson fermions,  which is not fully 
examined yet. Let us consider the unitarity in lattice QCD defined 
with Neuberger's (overlap) Dirac operator. 
For free theory, L\"uscher worked out 
the spectral representation of the fermion 
propagator: 
\[
\left\langle \psi(x) \bar \psi(y) \right\rangle_{x_0 > y_0}
=
\int_0^\infty d E  \int_{-\pi}^{\pi}
\frac{d^3 {\bf p}}{(2\pi)^3} \, \, 
\rho(E,{\bf p})  
\]
\begin{equation}
\qquad\qquad\qquad
 \times \, {\rm e}^{- E (x_0-y_0) + i \bm{p} (\bm{x - y})},
\end{equation}
where ($m_0=1$, $\hat p^2 =\sum_{k=1}^3 4 \sin^2 \frac{p_k}{2}$)
\begin{eqnarray}
\small
&&\rho(E,\bm{p})
= \small (\gamma_0 \sinh E - i\gamma_k \sin p_k ) \nonumber\\
&& \scriptstyle
\qquad
\times 
\left\{ 
\delta(E-\omega_{\bm{p}})
\theta\left( {\cosh E  -\frac{1}{2} {\hat p}^2 }\right)
\frac{\cosh E - \frac{1}{2} {\hat p}^2 }{\sinh 2E}
\right.\nonumber\\
&&\scriptstyle
\qquad
\left.
+ \frac{1}{2\pi}\theta(E-E_{\bm{p}})
\frac{
\left\{{\hat p}^2 \left(\cosh E- \cosh E_{\bm{p}}\right)\right\}^{1/2}}
%\scriptstyle 
%\left\{ {\hat p}^2 
%\left(\cosh E- \cosh E_\bm{p}\right) \right\}^{1/2}
%     }
     {
{\hat p}^2 \left(\cosh E - \cosh E_{\bm{p}}\right)
      +\left( \cosh E  - \frac{1}{2} {\hat p}^2 \right)^2 
     }
\right\} \nonumber
\end{eqnarray}
\begin{equation}
\scriptstyle
\cosh E_{\bm{p}}
= 1 + \frac{1}{\hat p^2}\left\{ 1+\frac{1}{4}
\sum_{k\not = l} \hat p_k^2 \hat p_l^2 \right\}, \ \ 
\sinh \omega_{\bm{p}} = \vert \sin^2 p_k \vert.
\end{equation}
This result shows that the spectral function is non-negative
\begin{equation}
  dE d^3 {\bf p} \, \zeta^\dagger \, 
\rho(E,{\bf p}) \, \zeta   \ge 0 
\end{equation}
and unitarity is maintained for any 
value of the lattice spacing $a$\cite{luscher-lecture}.

With gauge interaction, it is possible to examine the reflection 
positivity\cite{reflection-positivity} 
using the connection to domain wall fermion. 
For this purpose, we recall the fact that 
the partition function of the lattice fermion defined 
with Neuberger's (overlap) Dirac operator 
in a truncated approximation\cite{truncated-overlap}
can be expressed by the partition function of 
domain wall fermion. Namely, we have
\begin{equation}
\label{eq:DWF-partition-function-factorized}
\det D_N =
\frac{\det \left(D_{\rm 5w}-m_0\right)}
{{\det \left( D_{\rm 5w} -m_0\right) }_{[AP]} },
\end{equation}
where $D_N =\left(1 + \gamma_5 \tanh \left(a_5 N H \right)\right)/2$.
In the r.h.s. of 
Eq.~(\ref{eq:DWF-partition-function-factorized}),
the five-dimensional Wilson fermion subject to
the anti-periodic (AP) boundary condition in the 
fifth-dimension
is introduced\cite{vranas-pauli-villars}.
Moreover, {\it chiral invariant} observables in the original fermion system
can be expressed in terms of the boundary field variables
of the domain wall fermion\cite{kikukawa-noguchi}:
\[
  \langle {\cal O}\left[ U, (1-2 D_N)\psi, \bar \psi \right] \rangle^{(N)}
 =\langle {\cal O}\left[ U, q, \bar q\right] \rangle_{\rm DW}^{(N)}
\]
where $q(x) = \psi_L \left(x,-N+1 \right) + \psi_R \left(x,N \right)$.
This result follows from the identity,
\begin{equation}
\label{eq:light-fermion-propagator}
\left\langle q(x) \bar q(y) \right\rangle_F 
= 
{ D_N}^{-1}-\delta(x,y) .
\end{equation}

Since the r.h.s. of Eq.~(\ref{eq:DWF-partition-function-factorized}) 
may be expressed as
\begin{equation}
\longrightarrow 
\frac{ \det \left(D_{\rm 5w}-m_0\right) 
\cdot \det \left(D_{\rm 5w}-m_0\right)_{[AP]}}
{\det 
\left\{ \left( D_{\rm 5w} -m_0\right)^\dagger 
        \left( D_{\rm 5w} -m_0\right) \right\}_{[AP]} },
\end{equation}
the original fermion system is equivalent to the set of 
two five-dimensional Wilson-Dirac fermions (with Dirichlet b.c. and
anti-periodic b.c. in the fifth dimension, respectively)
and one complex four-component boson (with the anti-periodic b.c.).
\begin{eqnarray}
\label{eq:equivalent-DWF-action}
&& S_{\rm DW} = \sum_{x,t} \bar \psi (D_{\rm 5w}-m_0 ) \psi(x,t) 
\nonumber\\
&&  \quad \qquad 
+ \sum_{x,t}  \bar \psi^\prime (D_{\rm 5w}-m_0 ) \psi^\prime (x,t) \vert_{AP}
\nonumber\\
&& \quad \qquad
+ \sum_{x,t} \Phi^\dagger \vert D_{\rm 5w}-m_0 \vert^2 \Phi(x,t)
\vert_{AP}.
\end{eqnarray}
Then the question is reduced to the reflection positivity 
in the system given by the action Eq.~(\ref{eq:equivalent-DWF-action}) 
and for the observables written in terms of $U_k$, $q$ and
$\bar q$ in the gauge $U_4=1$.
Under the reflection symmetry transformation with respect to 
$x_4=1/2$ : $\theta (\bm{x},x_4) = (\bm{x},-x_4+1)$, $(\Psi=\psi, \psi^\prime)$
\begin{eqnarray*}
&& \theta U(x,x+\hat k) = U(\theta x, \theta (x+\hat k))\\
%&&  \theta \psi(x,t) = \bar \psi( \theta x , t) \gamma_4 , \ 
%    \theta \bar \psi(x,t) = \gamma_4  \psi(\theta x, t) \\
&&  \theta \Psi(x,t) = \bar \Psi( \theta x , t) \gamma_4 , \ 
    \theta \bar \Psi(x,t) = \gamma_4  \Psi(\theta x, t) \\
&& \theta \Phi(x,t) = \Phi^\dagger( \theta x , t) \gamma_4 \gamma_5 P
\ ( P: t \rightarrow -t +1)
\end{eqnarray*}
\begin{eqnarray*}
&& 
\theta 
\left( f_{\alpha_1,\cdot,\alpha_n}(U)\phi_{\alpha_1} 
\cdots \phi_{\alpha_n} \right)
\\
&& \qquad \quad
= f_{\alpha_1,\cdot,\alpha_n}^\ast(\theta U) 
\theta \phi_{\alpha_1} \cdots 
\theta \phi_{\alpha_n},
\end{eqnarray*}
$S_{\rm DW}$ can be written into the form
\begin{equation}
S_{\rm DW} = \theta B_+ + \sum_i (\theta A_+^i) A_+^i + B_+ ,
\end{equation}
where $A_+^i$ and $B_+$ are functions of the field variables
in the region $x_0 > 1/2$. Then we can show the positivity
\begin{equation}
  \langle \theta {\cal O}[U,q,\bar q]_+ 
\cdot {\cal O}[U,q,\bar q]_+ \rangle^{(N)} \ge 0 .
\end{equation}
The limit $N\rightarrow \infty$ is defined well as long as
the admissibility condition is assumed\cite{locality-in-dwf}. 
Therefore the above 
positivity implies the positivity in lattice QCD 
with Neuberger's (overlap) Dirac operator.
The analysis and discussion in more detail will be given 
elsewhere\cite{kikukawa-reflection-positivity}.

The Hamiltonian approach is a possible direction to
the formulation of lattice QCD with manifest 
chiral symmetry and unitarity. 
Creutz, Horvath and Neuberger has 
constructed such a Hamiltonian based on the overlap 
construction\cite{hamiltonian-overlap}.
It is defined through the three-dimensional Wilson-Dirac opertor 
$X = D_{\rm 3w}- m_0 (0 < m_0 < 2)$ as 
$\hat H_F = \sum_x \psi^\dagger(x) H_F \psi(x)$ where 
\begin{equation}
H_F=\gamma_0 \left(1+X /\sqrt{X^\dagger X}\right)\equiv\gamma_0 D .   
\end{equation}
Because of the property of $D$ in three dimensions,
$D^\dagger = \gamma_5 D \gamma_5 = \gamma_0 D \gamma_0$, 
there are two kinds of the Ginsparg-Wilson relations
with $\gamma_5$ and $\gamma_0$, respectively.
The chiral charge can be introduced as 
$\hat Q_5 = \sum_x \psi^\dagger(x) Q_5 \psi(x)$ where
\begin{equation}
Q_5 = \gamma_5 \left( 1-\frac{1}{2}D\right)
\end{equation}
and it commutes with the Hamiltonian $\hat H_F$. But it
does not commute with the electric term in the Hamiltonian of the 
gauge field, reflecting chiral anomaly. 
$H_F$ and $Q_5$ have a correlation
\begin{equation}
  Q_5^2 + \left( \frac{H_F}{2} \right)^2 = 1,
\end{equation}
which leads to an interesting interpretation of 
the chiral properties of the low-lying
eigenmodes of $H_F$.
This approach should have physical applications
in QCD and also in electroweak theory.

\subsection{Algebraic generalization}

It is an important and challenging issue to 
seek new lattice Dirac operators with better chiral and locality properties. 
Fujikawa and Ishibashi has examined this possibility
through an algebraic generalization of the Ginsparg-Wilson 
relation\cite{algebraic-generalization-GW,fujikawa-ishibashi}. 
They consider the following relation for a lattice Dirac 
operator:
\begin{equation}
\label{eq:algebraic-generalization}
 \gamma_5 \left( \gamma_5 D \right) 
+\left( \gamma_5 D \right) \gamma_5  = 
2 a^{2k+1} \left(\gamma_5 D \right)^{2k+2} 
\end{equation}
for $k=0,1,\cdots$. The case $k=0$ corresponds to 
the usual Ginsparg-Wilson relation. This generalization is intended 
so that the index relation should be maintained:
\begin{equation}
{\rm Tr} \Gamma_5 = n_+ - n_- ,  \quad 
\Gamma_5 \equiv \gamma_5 - \left(\gamma_5 a D\right)^{2k+1}.
\end{equation}
The explicit solution of the relation 
Eq.~(\ref{eq:algebraic-generalization}) has been constructed 
by noting the relation
\begin{equation}
\gamma_5 H^{2k+1} + H^{2k+1} \gamma_5 = 2 H^{2(2k+1)}
\end{equation}
for $H= \gamma_5 D$ and applying the overlap construction to $H^{2k+1}$.
The consistency of free theory, chiral anomaly and locality has been 
checked\cite{fujikawa-ishibashi}. The specturm of the class of 
the lattice Dirac operators has been examined numerically by
Chiu\cite{algebraic-generalization-chiu}.
Unfortunately, it turned out that the locality property becomes worse
for larger values of $k$. Therefore this approach cannot be used 
to improve the locality property, although the algebraic structure
is interesting. 
It is desirable to have the systematic understanding about
the conditions which 
determines the locality properties of a lattice Dirac operator.

\subsection{Odd dimensions}

Bietenholz and Nishimura has argued that 
the Ginsparg-Wilson relation is also useful as a condition
for lattice Dirac operator of massless fermions in odd 
dimensions\cite{bietenholz-nishimura}.
They adopt the following relation for the three-dimensional
Dirac operator:
\begin{equation}
\label{eq:GW-rel-odd-dim}
  D + D^\dagger =  D^\dagger D ,
\end{equation}
which solution may be written in general as 
$D =\left( 1 - V \right)$,  $V^\dagger V = 1$.
The fermion action defined with such a Dirac operator
is then invariant under the following parity transformation: 
$R: x \rightarrow -x$,
\begin{eqnarray}
%&&  R: x \rightarrow -x  \\
&& U(x,\mu) \rightarrow U(x,\mu)^P = U^\dagger(-x-\mu,\mu) \\
&& \psi(x) \rightarrow i R V \psi(x) \ ; \quad 
  \bar \psi(x) \rightarrow i \bar \psi(x) R 
\end{eqnarray}
An interesting observation made here is that 
by this discrete transformation, the fermion measure 
transforms non-trivially,
\begin{equation}
  d \psi d \bar \psi \rightarrow (\det V)^{-1} d \psi d \bar \psi
\end{equation}
and the Jacobian can be regarded as the parity anomaly.

The explicit solution of the relation Eq.~(\ref{eq:GW-rel-odd-dim})
is again provided by the vacuum overlap 
formalism\cite{narayanan-nishimura,overlap-D-odd-dim}.
It has been known that in odd dimensions, 
it is possible to choose a phase 
so that the overlap formula is given by
a determinant of a gauge-covariant operator, $D_{\rm ov}$ \cite{overlap-D-odd-dim}.
%\begin{equation}
%  D_{\rm ov}= \frac{1}{2}\left(1+ X \frac{1}{
%\sqrt{X^\dagger X}} \right),
%%  
%\end{equation}
%where $X=D_{\rm 3w}-m_0 (0 < m_0 < 2)$.
The complex phase of the determinant of $D_{\rm ov}$ is nothing
but the complex phase of the determinant of the three-dimensional
Wilson-Dirac operator
%\begin{equation}
%  {\rm Im} \ln \det D_{\rm ov} 
% ={\rm Im} \ln \det \left(D_{\rm 3w}-m_0\right)
%\end{equation}
and it reproduces the Chern-Simons term
in the continuum limit\cite{so,coste-luscher}. 

Form the above point of view based on the Ginsparg-Wilson relation, 
this complex phase is just identified as the parity anomaly
from the functional measure
\begin{equation}
  {\rm Im} \ln \det V 
 ={\rm Im} \ln \det \left(D_{\rm 3w}-m_0\right) 
\end{equation}
and it provides a reasonable definition of the lattice Chern-Simons
term\cite{bietenholz-nishimura}. 
In five-dimensions, this picture
fits well to the structure of the descent relation 
over four-, five-, and six-dimensional lattices discussed 
in section~\ref{sec:chiral-gauge-theory}.
The universality classes of the three-dimensional Dirac fermion
\cite{coste-luscher} has also been examined. % from this point of view.

\section{Conclusion}
Core problems still remain.
In chiral theories, the construction of non-abelian chiral gauge theories
is still an open question. We need to know how to treat
the space of the admissible lattice gauge fields. 
In vector-like theories, we should have better understanding 
of the behavior of the Ginsparg-Wilson fermions
without the admissibility condition (cf. \cite{brower-et-al}). 
In particular for Neuberger's (overlap) Dirac fermion,
the behavior of low-lying eigenvalues of hermitian Wilson-Dirac 
operator should be understood 
(cf. \cite{narayanan-neuberger-berruto,taniguchi-aoki-4d-eigenvalue,gattringer-et-al}.)

For Neuberger's (overlap) Dirac fermion,
the relation to domain wall fermion 
is quite useful in theoretical investigations.
%hopefully as explained 
%to some extent in this article.
In numerical investigations, on the other hand, 
it seems relatively simpler to treat 
the four-dimensional Dirac operator directly. Still the five-dimensional 
point of view would be useful to get insights into the problem
we face (cf. \cite{borici-continued-fraction,staggered-dwf}). 

%The recent re-discovery of the Ginsparg-Wilson relation and the 
%realization of exact chiral symmetry on the lattice
%are quite developments
%from more general point of view 
%of quantum field theory. 
%To extend this nice idea to 
%other aspects of quantum field theory would be another
%big challenge. 
%%Hope is that 
%%such studies would constribute to the %researches towards the
%%understanding 
%%of the dynamics of electroweak gauge symmetry breaking
%%and other fundamental problems in particle physics. 

\section*{Acknowledgments}
The author would like to thank M.~L\"uscher, P.~Hern\'andez, H.~Suzuki
and Y.~Nakayama for valuable discussions.
He is also grateful to T.-W.~Chiu, S.~Aoki, Y.~Taniguchi,
W. Bietenholz and M.~Mueller-Preussker.
Y.K. is supported in part by Grant-in-Aid 
for Scientific Research of Ministry of Education (\#10740116).


\begin{thebibliography}{99}
\bibitem{ginsparg-wilson-rel}
P.~H.~Ginsparg and K. G. Wilson, Phys. Rev. D25 (1982) 2649. 

\bibitem{fp-action-D}
P.~Hasenfratz, Nucl. Phys. B(Proc. Suppl.) 63 (1998) 53. 

\bibitem{overlap-D}
H.~Neuberger, Nucl. Phys. B417 (1998) 141;
Phys. Lett. B427 (1998) 353.

\bibitem{locality-of-overlap-D}
P.~Hern\'andez, K.~Jansen and M.~L\"uscher, Nucl. Phys. B552 (1999) 363.

\bibitem{exact-chiral-symmetry}
M. L\"uscher, Phys. Lett. B428 (1998) 342.

\bibitem{kikukawa-yamada}
Y.~Kikukawa and A.~Yamada, Phys. Lett. B448 (1999) 265.

\bibitem{adams}
D.H.~Adams, hep-lat/9812019.

\bibitem{fujikawa}
K.~Fujikawa, Nucl. Phys. B546 (1999) 480.

\bibitem{suzuki}
H.~Suzuki, Prog. Theor. Phys. 102 (1999) 141.

\bibitem{chiu}
T.-W.~Chiu, Phys. Lett. B445 (1999) 371; 
T.-W.~Chiu and T.-H.~Hsieh, hep-lat/9901011.

\bibitem{index-theorem-on-lattice}
P.~Hasenfratz, V.~Laliena, F.~Niedermayer, 
Phys. Lett. B427 (1998) 125.

\bibitem{overlap}
R.~Narayanan and H.~Neuberger, 
Nucl. Phys. B412 (1994) 574;
Phys. Rev. Lett. 71 (1993) 3251;
Nucl. Phys. B443 (1995) 305.

\bibitem{abelian-chiral-gauge-theory}
M.~L\"uscher, Nucl. Phys. B549 (1999) 295.

\bibitem{nonabelian-chiral-gauge-theory}
M.~L\"uscher, Nucl. Phys. B568 (2000) 162.

\bibitem{effective-action-abelian-chiral}
H.~Suzuki, Prog. Theor. Phys. 101 (1999) 1147.

\bibitem{neuberger-bound}
H.~Neuberger, Phys. Rev. D61 (2000) 085015.

\bibitem{Pilar-review}
P.~Hern\'andez, these proceedings.

\bibitem{domain-wall-fermion}
D.~B.~Kaplan, Phys. Lett. B288 (1992) 342. 

\bibitem{boundary-fermion}
Y.~Shamir, Nucl. Phys. B406 (1993) 90. 

\bibitem{boundary-fermion-QCD}
V.~Furman, Y.~Shamir, Nucl. Phys. B439 (1995) 54.

\bibitem{blum-soni}
T.~Blum and A.~Soni, Phys. Rev. D56 (1997) 174;
Phys. Rev. Lett. 79 (1997) 3595.

%
%\bibitem{aoki-taniguchi}
%S.~Aoki, Y.~Taniguchi, Phys. Rev. D59 (1999) 054510.
%
%\bibitem{kikukawa-neuberger-yamada}
%Y.~Kikukawa, H.~Neuberger, A.~Yamada, Nucl. Phys. B526 (1998) 572.
%

\bibitem{aoyama-kikukawa} 
T.~Aoyama, Y.~Kikukawa, hep-lat/9905003.
%"A lattice implementation of the $\eta$-invariant 
%and effective action for chiral fermions on the lattice", 

\bibitem{chiral-domain-wall-fermion}
Y.~Kikukawa, hep-lat/0105032.

\bibitem{so}
H.~So, Prog. Theor. Phys. 73 (1985) 582; 74 (1985) 585.

\bibitem{coste-luscher}
A.~Coste and M. L\"uscher, Nucl. Phys. B323 (1989) 631.

\bibitem{topology-on-the-lattice}
M.~L\"uscher, Nucl. Phys. B538 (1999) 515.

\bibitem{noncomutative-brs-cohomology}
T.~Fujiwara, H.~Suzuki, K.~Wu, 
Phys. Lett. B463 (1999) 63; Nucl. Phys. B569 (2000) 643.

\bibitem{perturbation-theory}
M.~L\"uscher, JHEP 0006 (2000) 028. 

\bibitem{cohomology-non-abelian}
H.~Suzuki, Nucl. Phys. B585 (2000) 471.
%; H.~Igarashi, K.~Okuyama, H.~Suzuki, {\tt hep-lat/0012018}.

\bibitem{truncated-overlap}
H.~Neuberger, Phys. Rev. D57 (1998) 5417.

\bibitem{luscher-dwf}
M.~L\"uscher, private communication.

\bibitem{borici}
A.~Bori\c ci, Nucl. Phys. Proc. Suppl. 83 (2000) 771; hep-lat/9912040. 

\bibitem{adiabatic-phase-chioce}
H.~Neuberger, Phys. Rev. D63 (2001) 014503.

\bibitem{geometrical-aspect}
H.~Neuberger, Phys. Rev. D59 (1999) 085006.

%\bibitem{descent-relation-stora}
%R.~Stora, Constinuum gauge theories, in: New developments
%in quantum field theory and statistical mechanics 
%(Car\`eges 1976), eds. M.~L\'evy and P.~Mitter
%(Plenum Press, New York, 1977);
%Algebraic structure and topological origin of anomalies, in:
%Progress in gauge field theory (Car\`eges 1977), 
%eds. G.`t Hooft et al. (Plenum Press, New York, 1984).

%\bibitem{descent-relation-zumino}
%B.~Zumino, Chiral anomalies and differential geometry, in: 
%Relativity, groups and topology (Les Houches 1983), 
%eds, B.S.~DeWitt and R.~Stora (North-Holland, Amsterdam, 1984).

\bibitem{kikukawa-nakayama-suzuki}
Y.~Kikukawa, Y.~Nakayama and H.~Suzuki, 
these proceedings.

\bibitem{ML-topology}
M. L\"uscher,
%\emph{Topology of lattice gauge fields},
Comm. Math. Phys. 85 (1982) 39.

\bibitem{real-rep-suzuki}
H.~Suzuki, JHEP 0010 (2000) 039.

\bibitem{neuberger-su2}
H.~Neuberger, Phys. Lett. B437 (1998) 117.


\bibitem{OB-IC}
O. B\"ar, I. Campos,
%%\emph{Global anomalies in chiral lattice gauge theory},
%\npps{83}{2000}{594};
%\emph{Global anomalies in chiral gauge theories on the lattice},
Nucl. Phys. B581 (2000) 499.

\bibitem{YK-YN-anomaly}
Y. Kikukawa, Y. Nakayama,
%\emph{Gauge anomaly cancellation in SU(2)$_L\times$U(1)$_Y$ electroweak theory on the lattice}, 
Nucl. Phys. B597 (2001) 519.

\bibitem{luscher-lecture}
M.~L\"uscher, 
%Lectures given at the International School of 
%Subnuclear Physics, Erice, 27 August - 5 September 2000, 
hep-th/0102028.

\bibitem{reflection-positivity}
K.~Osterwalder and E.~Seiler, Ann. Phys. (NY) 110 (1978) 440;
M.~L\"uscher, Commun. Math. Phys. 54 (1977) 283;
P.~Menotti and A.~Pelissetto, Commun. Math. Phys. 113 (1987) 369.

%\bibitem{osterwalder-seiler}
%\bibitem{luscher-transfer-matrix}
%\bibitem{menotti-pelisseto}

\bibitem{vranas-pauli-villars}
P.~Vranas, Phys. Rev. D57 (1998) 1415.

\bibitem{kikukawa-noguchi}
Y.~Kikukawa, T.~Noguchi, hep-lat/9902022.
%\\
%``Low energy effective action of domain-wall fermion 
%and the Ginsparg-Wilson relation'', 


\bibitem{locality-in-dwf}
Y.~Kikukawa, Nucl. Phys. B584 (2000) 511.

\bibitem{kikukawa-reflection-positivity}
Y.~Kikukawa, in preparation.

%\bibitem{holvath-et-al}
% I.~Horvath, C.T.~Balwe, R.~Mendris, 
%Nucl. Phys. B599 (2001) 283-304.

\bibitem{hamiltonian-overlap}
M,~Creutz, I.~Horvath and H.~Neuberger, 
these proceedings.

\bibitem{algebraic-generalization-GW}
K.~Fujikawa, Nucl. Phys. B589 (2000) 48.

\bibitem{fujikawa-ishibashi}
K.~Fujikawa, M.~Ishibashi, Nucl. Phys. B587 (2000) 419;
Nucl. Phys. B605 (2001) 365; these proceedings.

\bibitem{algebraic-generalization-chiu}
T.-W.~Chiu, Nucl. Phys. Proc. Suppl. 94 (2001) 733.

\bibitem{bietenholz-nishimura}
W.~Bietenholz and J.~Nishimura, JHEP 0107 (2001) 015.

\bibitem{narayanan-nishimura}
R.~Narayanan, J.~Nishimura, Nucl. Phys. B508 (1997) 371.

\bibitem{overlap-D-odd-dim}
Y.~Kikukawa and H.~Neuberger, Nucl. Phys. B513 (1998) 735.

\bibitem{brower-et-al}
F.~Berruto, R.C.~Brower, B.~Svetitsky, 
Phys. Rev. D64 (2001) 114504.

\bibitem{narayanan-neuberger-berruto}
F.~Berruto, R.~Narayanan, H.~Neuberger, 
Phys. Lett. B489 (2000) 243.

\bibitem{taniguchi-aoki-4d-eigenvalue}
S.~Aoki, Y.~Taniguchi, hep-lat/0109022; 
these proceedings. %(hep-lat/0110126).

%\bibitem{izubuchi}

\bibitem{gattringer-et-al}
C.~Gattringer et al., hep-lat/0108001. 

\bibitem{borici-continued-fraction}
A.~Borici et al., %A.D.~Kennedy, B.J.~Pendleton, U.~Wenger, 
these proceedings. %(hep-lat/0110070).

\bibitem{staggered-dwf}
G.~Fleming, P.~Vranas, 
these proceedings. %(hep-lat/0110172).

%\bibitem{nielsen-ninomiya}
%N.B.~Neilsen and M.~Ninomiya, PL B105 (1981) 219; Nucl. Phys. B185
%(1981) 20 [E: B195 (1982) 541]; {\it ibid} B193 (1981) 173. 
%Friedan, Commun. Math. Phys. 85 (1982) 481.
%
%\bibitem{Wilson-mass}
%K.G.~Wilson, Phys. Rev. D10 (1974) 2445; {\it in} New phenomena in 
%subnuclear physics, ed. A.~Zichichi (Plenum, New York, 1977)
%(Erice, 1975)
%
%\bibitem{eta-invariant}
%L. Alvarez-Gaum\'e, S. Della Pietra and V. Della Pietra, 
%Phys. Lett. 166B (1986) 177;
%Commun. Math. Phys. 109, (1987) 691-700.
%
%\bibitem{ball-osbor}
%R.D.~Ball and H.~Osborn, Phys. Lett. B165 (1985) 410;
%Nucl. Phys. B263 (1986) 243. 
%R.D.~Ball, Phys. Lett. B171 (1986) 435; Phys. Rept. 182 (1989) 1.   
%
%%\bibitem{atiyah-padoti-singer}
%%M.F.~Atiya, V.K.~Patodi and I.M.~Singer, 
%%Math. Proc. Camb. Phil. Soc. 77 (1975) 43; 
%%78 (1975) 405; 79 (1976) 71.
%
%\bibitem{domain-wall-ferimon-and-eta-invariant}
%D.~B.~Kaplan and  M.~Schmaltz, Phys. Lett. B368 (1996) 44.
%


\end{thebibliography}
\end{document}